\documentclass[twocolumn,english,aps,superscriptaddress,prl]{revtex4-2}
\usepackage{color}
\usepackage{amsmath}
\usepackage{graphicx}
\usepackage{amssymb}
\usepackage{hyperref}

\usepackage{bm}

\usepackage{ulem}

\newcommand{\w}{\omega}

\newcommand{\KK}{K}   % Kitaev coupling
\newcommand{\JK}{J_{\rm K}}   % Kondo coupling
\newcommand{\TK}{T_{\rm K}}   % Kondo temperature
\newcommand{\nc}{n_c}   % Occupation number

\newcommand{\Ztwo}{\mathbb{Z}_2}

\newcommand{\SUtwo}{\mathrm{SU(2)}}

\newcommand{\HH}{\mathcal{H}}
\newcommand{\Ht}{\mathcal{H}_t}
\newcommand{\HK}{\mathcal{H}_K}
\newcommand{\HJ}{\mathcal{H}_J}

\newcommand{\Hint}{\mathcal{H}_{\rm int}}

\newcommand{\pd}{{\phantom{\dagger}}}

\graphicspath{{./}{./figs/}}

\begin{document}

\title{
Fractionalized Superconductivity Mediated by Majorana Fermions in the Kitaev-Kondo Lattice
}

\author{Matthew Bunney}
\affiliation{School of Physics, University of Melbourne, Parkville, VIC 3010, Australia}
\affiliation{Institute for Theoretical Solid State Physics, RWTH Aachen University, 52062 Aachen, Germany}
\author{Urban F.\ P.\ Seifert}
\affiliation{Institut f\"ur Theoretische Physik, Universit\"at zu K\"oln, Z\"ulpicher Str. 77a, 50937 K\"oln, Germany}
\author{Stephan Rachel}
\affiliation{School of Physics, University of Melbourne, Parkville, VIC 3010, Australia}
\author{Matthias Vojta}
\affiliation{Institut f\"ur Theoretische Physik and W\"urzburg-Dresden Cluster of Excellence ct.qmat, Technische Universit\"at Dresden, 01062 Dresden, Germany}

%%%%%%%%%%%%%%%%%%%%%%%%%%%%%%%%%%%%%%%%%%%%%%%%%%%%%%%%%%%%%%%%%%%%%%%

\begin{abstract}
Superconductivity usually emerges from a metallic normal state which follows the Fermi-liquid paradigm. If, in contrast, the normal state is a fractionalized non-Fermi liquid, then pairing may either eliminate fractionalization via a Higgs-type mechanism leading to a conventional superconducting state, or pairing can occur in the presence of fractionalization. Here we discuss a simple model for the latter case: Using a combination of perturbation theory and functional renormalization group, we show that the Kitaev--Kondo lattice model displays a fractionalized superconducting phase at weak Kondo coupling. This phase is characterized by Cooper pairing of conventional electronic quasiparticles, coexisting with a spin-liquid background and topological order. Depending on the sign of the Kitaev coupling, we find the pairing to be either of chiral $d$-wave or $p$-wave type for extended doping regions around the van-Hove filling. We discuss applications and extensions.
\end{abstract}

\date{\today}

\maketitle
%%%%%%%%%%%%%%%%%%%%%%%%%%%%%%%%%%%%%%%%%%%%%%%%%%%%%%%%%%%%%%%%%%%%%%%

%\section{Introduction}
Fractionalized phases of matter in space dimensions $d\geq 2$ attract widespread interest in condensed matter research: They are characterized by emergent gauge fields, topological order, and fractionalized excitations which constitute novel quasiparticles \cite{wen2017,balents2010}. Fractionalization is often discussed for bulk insulators such as gapped spin liquids and fractional Quantum-Hall states.
Significantly less work has been devoted to fractionalized metals and superconductors, mainly because the presence of bulk charge carriers renders both conceptual and experimental distinctions between fractionalized and conventional phases more subtle than in the simpler insulating cases.

One class of fractionalized metals whose existence can be rigorously established theoretically is that of fractionalized Fermi liquids (FL$^\ast$) \cite{senthil2003,senthil2004}: A two-band model where electrons in one strongly correlated band form a fractionalized spin liquid of local moments, e.g. because of frustration, and are weakly exchange-coupled to a second band of weakly correlated electrons realizes an FL$^{\ast}$ phase. The FL$^{\ast}$ phase is characterized by the coexistence of fractionalized spin excitations, and their associated gauge field, as well as charge carriers with charge $|e|$ and spin $1/2$ whose Fermi volume violates Luttinger's theorem \cite{oshikawa2000}. A stronger coupling between the two bands leads to Kondo screening of the local moments, resulting in a more conventional heavy Fermi-liquid (FL) phase. Consequently, the transition between FL and FL$^{\ast}$ is an instance of a Kondo-breakdown quantum phase transition \cite{coleman2001,si2001,senthil2003,lohneysen2007,vojta2010}, accompanied by a jump in the Fermi volume but no spontaneous symmetry breaking.

The focus of this paper is on constructing a superconducting cousin of the FL$^\ast$ phase, i.e., a fractionalized superconductor (SC$^\ast$), where superconductivity coexists with topological order. In fact, this has been hypothesized early on to exist in underdoped cuprate superconductors \cite{senthil2000}. More recently, it has been discussed \cite{christos2024} in scenarios of the cuprate pseudogap representing a fractionalized Fermi liquid \cite{qi2010,moon2011,mei2012}. A transition between fractionalized and conventional superconductivity (SC) has also been proposed to occur in an iron-based superconductor \cite{chowdhury2013}.

\begin{figure}[t!]
\centering
\includegraphics[width=\columnwidth]{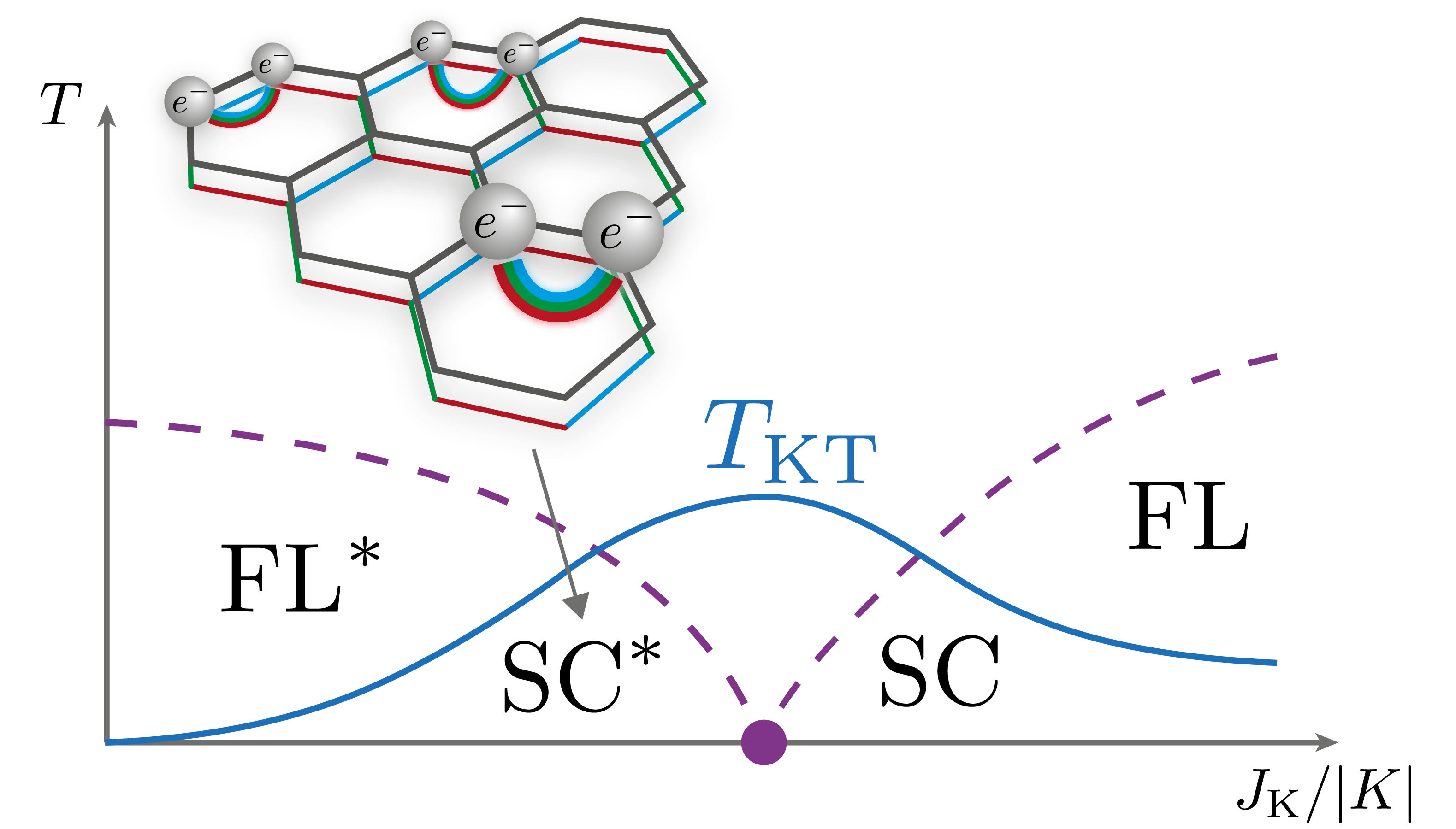}
\caption{Schematic phase diagram of the Kitaev--Kondo lattice as function of temperature and Kondo screening strength. While the FL$^\ast$, FL, and SC phases are captured by mean-field theory, the fractionalized superconductor SC$^\ast$ discussed in this paper is not. A quantum critical point separates the fractionalized from the non-fractionalized phases. In two space dimensions, the superconducting phases are bounded by a line of Kosterlitz-Thouless transitions at $T_{\rm KT}$.
}
\label{fig:pd}
\end{figure}

In this paper we discuss and solve a tractable model of fractionalized superconductivity. We consider the Kitaev--Kondo lattice model, Fig.~\ref{fig:pd}, constructed from the honeycomb-lattice spin-liquid model of Kitaev \cite{kitaev2006} and a band of conduction electrons. This model has been shown to display a FL$^\ast$--FL transition which -- at the mean-field level -- is masked by non-trivial but non-fractionalized superconductivity at intermediate Kondo coupling \cite{seifert2018,choi2018}.
Here we focus on the fractionalized weak-coupling regime where we can perturbatively derive an effective model for conduction-electron interactions mediated by the Majorana fermions of the spin liquid.
We solve this model using the functional renormalization group (FRG) technique \cite{metzner2012,platt2013} and find stable superconductivity of either chiral $d$-wave ($p$-wave) type for ferromagnetic (antiferromagnetic) Kitaev coupling, respectively, over a range of band fillings. These pairing states coexist with the fractionalized spin liquid in the local-moment sector, rendering this SC$^\ast$ phase a natural descendant of the model's FL$^\ast$ phase.

\iffalse
The remainder of the paper is organized as follows: In Sec.~\ref{sec:model} we review the Kitaev Kondo lattice model and the results of its mean-field treatment. We then derive an effective conduction-electron model in Sec.~\ref{sec:eff} which is analyzed using FRG in Sec.~\ref{sec:fRG}. The superconducting instabilities identified by FRG, and the resulting SC$^\ast$ phases, are then investigated using mean-field theory in Sec.~\ref{sec:mft}. Finally, we briefly discuss the Kondo-driven quantum phase transition between SC$^\ast$ and a more conventional SC phase. An outlook closes the paper.
\fi

%%%%%%%%%%%%%%%%%%%%%%%%%%%%%%%%%%%%%%%%%%%%%%%%%%%%%%%%%%%%%%%%%%%%%%%

{\it Kitaev--Kondo lattice model.---}
\label{sec:model}
We consider a Kondo--lattice model on a two-dimensional honeycomb lattice, with the key ingredient being Kitaev interactions between the local moments\,\cite{kitaev2006}. The total Hamiltonian is $\HH=\Ht+\HK+\HJ$ with
\begin{align}
\Ht &= - t \sum_{\langle ij\rangle\sigma} (c_{i\sigma}^\dagger c_{j\sigma} + {\rm H.c.}),  \label{ht} \\
%\end{align}
%\begin{align}
\HK & = -  \sum_{\langle ij\rangle_\alpha} \KK^\alpha S_i^\alpha S_j^\alpha, \label{hK}  \\%[6pt]
\HJ & =  \sum_{i\alpha} \JK^\alpha s_i^\alpha S_i^\alpha \label{hJ}
\end{align}
in standard notation, with $S_i^\alpha$ representing the Kitaev spin at site $i$ and $s^\alpha_i = c_{i\sigma}^\dagger \tau^\alpha_{\sigma\sigma'} c_{i\sigma'} / 2$ the conduction-electron spin operator with Pauli matrices $\tau^\alpha$.
The first term represents the conduction-electron kinetic energy, the second the Kitaev coupling among the spin-$1/2$ local moments, with $\langle ij\rangle_\alpha$ an $\alpha$ bond on the lattice ($\alpha=x,y,z$), and the last the local Kondo coupling.
A chemical potential $\mu$ is applied to the conduction electrons to control their filling,
$\nc = \frac{1}{2N}\sum_{i\sigma} c_{i\sigma}^\dagger c_{i\sigma}$,
where $N$ is the number of unit cells. Half filling corresponds to $\nc=1$.
We will concentrate on $\nc \geq 1$; phases for $\nc \leq 1$ are related by particle--hole symmetry.

The symmetry properties of the model $\HH$, with isotropic Kitaev and Kondo couplings, are dictated by the symmetries of the Kitaev model $\HK$.
Its spin structure breaks the continuous $\SUtwo$ spin symmetry down to $\Ztwo\times\Ztwo$ \cite{baskaran08,you2012,liu24}, and combinations of spin and lattice transformations are further discrete symmetries\,\cite{you2012,rousochatzakis2015}.

The Kitaev model $\HK$ alone describes an exactly solvable $\Ztwo$ spin liquid \cite{kitaev2006}. Its degrees of freedom are itinerant ``matter'' Majorana fermions and static $\Ztwo$ gauge fluxes. Here we focus on  isotropic Kitaev couplings, $\KK^\alpha \equiv \KK$, for which the matter-Majorana spectrum is gapless, and on isotropic Kondo couplings, $\JK^\alpha \equiv \JK$.
Upon adding the Kondo coupling, the $\Ztwo$ fluxes are no longer conserved, and the solubility is spoiled.

We quickly summarize the results of Refs.~\onlinecite{seifert2018,choi2018} where the Kitaev--Kondo lattice model $\HH$ has been studied using perturbative and mean-field techniques. The model hosts a $\Ztwo$ fractionalized Fermi liquid \cite{senthil2003} for $\JK\ll\KK,t$ because the Kitaev spin liquid is stable against a small coupling to conduction electrons. Conversely, the model realizes a conventional heavy Fermi liquid for $\KK \ll \JK \sim t$ (or $\KK \ll \TK$ where $\TK$ is the Kondo temperature) with Kondo screening of the local moments.
According to the mean-field results, the transition between these two symmetric metallic phases is masked by superconductivity, in accordance with general considerations \cite{senthil2003}. At the mean-field level, superconductivity sets in concomitantly with Kondo screening, hence the superconducting phase is non-fractionalized (but may have non-trivial topological band invariants). The precise character of the superconducting phases depends on the mean-field scheme and differs between Refs.~\onlinecite{seifert2018} and \onlinecite{choi2018}. We note that the Kitaev--Kondo lattice has also been utilized to describe heterostructures composed of graphene and $\alpha$-RuCl$_3$ \cite{leeb2021}.

%%%%%%%%%%%%%%%%%%%%%%%%%%%%%%%%%%%%%%%%%%%%%%%%%%%%%%%%%%%%%%%%%%%%%%%

{\it Effective model at small Kondo coupling.---}
\label{sec:eff}
In the limit $\JK \ll \KK,t$ the topological character of the FL$^\ast$ phase is protected by the vison gap of the Kitaev spin liquid. Then, a perturbative treatment of $\JK$ is permissible. Since our goal is to describe electronic pairing, we choose to integrate out the local-moment degrees of freedom. This generates an effective spin--spin interaction among the conduction electrons. In second-order perturbation theory, this reads
\begin{align}
 \mathcal{S}_{\rm int} &= \sum_{ij} s_i^\alpha(\w) V^{\alpha\beta}(i,j;\w) s_j^\beta(\w)\ , \notag \\
 V^{\alpha\beta}(i,j;\w) &= \JK^2 \, \chi_K^{\alpha\beta}(i,j;\w)\,.
\end{align}
The quantity $\chi_K$ is the spin susceptibility of the Kitaev spin liquid; it has the remarkable property that it is non-zero only if the two sites $i,j$ are identical or nearest neighbors \cite{baskaran2007,knolle2014}.

%%%%%%%%%%%%%%%%%%%%%%%%%%%%%%%%%%%%%%%%%%%%%%%%%%%%%%%%%%%%%%%%%%%%%%%
\begin{figure*}[t!]
\centering
\includegraphics[width=\textwidth]{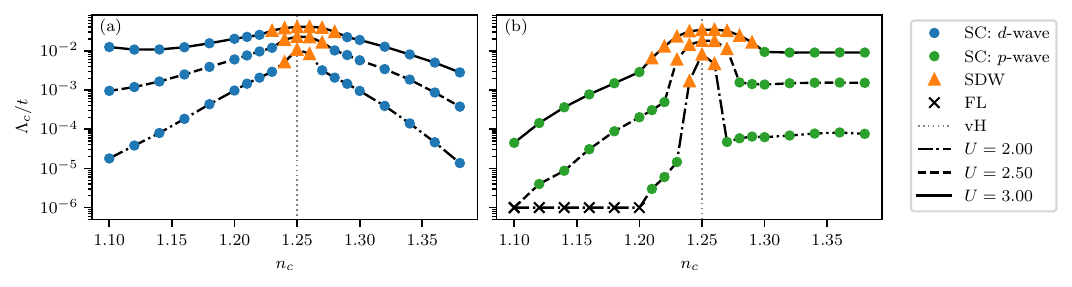}
\caption{FRG phase diagrams of the (a) AFM and (b) FM Hubbard--Kitaev model as a function of critical scale $\Lambda_c$ vs.\ filling $\nc$.
A magnetically ordered phase (SDW) at and near van-Hove filling $n_{\rm vH}=5/4$ is surrounded on both sides by a chiral $d$-wave, and $p$-wave superconductor, respectively. Shown are results for $U/t=2$, 2.5 and 3. The crosses indicate the absence of any divergence, i.e., the Fermi liquid prevails down to scales of $\Lambda=10^{-6}$.
}
\label{fig:FRG}
\end{figure*}
%%%%%%%%%%%%%%%%%%%%%%%%%%%%%%%%%%%%%%%%%%%%%%%%%%%%%%%%%%%%%%%%%%%%%%%

To proceed, we neglect the frequency dependence of $\chi_K$, i.e., we replace $\chi_K^{\alpha\beta}(i,j;\w)$ by its static value $\chi_K^{\alpha\beta}(i,j)$; this approximation is justified at weak coupling provided that this frequency dependence is regular. Then, the susceptibility is parameterized by only two numbers, $\chi_{0}$ and $\chi_{1}$: For $i=j$ we have $\chi_K^{\alpha\beta}(i,i)=\delta^{\alpha\beta} \chi_0$, while for nearest neighbors $(i,j)$ on a $\gamma$ bond $\chi_K^{\alpha\beta}(i,j)=\delta^{\alpha\gamma} \delta^{\beta\gamma} \chi_1$.
The effective interaction for the conduction electrons thus reduces to
\begin{align}
\Hint = \JK^2 \Big( \chi_0 \sum_i \vec{s}_i^2 + \chi_1 \sum_{\langle ij\rangle_\alpha} s_i^\alpha s_j^\alpha \Big)\ .
\end{align}
We note that $\chi_0<0$ while $\chi_1<0$ ($\chi_1>0$) for ferromagnetic (antiferromagnetic) Kitaev interaction, respectively.
In this interaction, the coupling constants of the model $\HH$ only enter in the combination $\JK^2/|K|$ since $\chi_{0,1}\propto 1/|K|$. We therefore introduce the new interaction parameter $U$ and make use of the identity $\vec{s}_i^2= -3/2 (n_{i\uparrow}-1/2)(n_{i\downarrow}-1/2) + 3/8$ to obtain
\begin{align}
\label{eff_int}
\Hint = U \bigg[ \sum_i n_{i\uparrow} n_{i\downarrow} + \kappa \sum_{\langle ij\rangle_\alpha} s_i^\alpha s_j^\alpha \bigg]
\end{align}
up to a constant, with $U=|\chi_0|J_K^2>0$ the effective interaction scale. The ratio of the on-site Hubbard interaction and the Kitaev-type spin interaction is given by $\kappa = \mp 2/3 |\chi_1/\chi_0|$ where the upper (lower) sign is for ferromagnetic (antiferromagnetic) Kitaev interaction, respectively.
We compute $\chi_{0,1}$ by a Kramers-Kronig transformation of $\chi''$ from Ref.~\onlinecite{knolle2014} and obtain $|\chi_1/\chi_0|=0.8897$. Therefore $\kappa=\mp 0.5931$\,\cite{footnote}.

The model $\Ht+\Hint$, below referred to as AFM/FM Hubbard--Kitaev model, describes the conduction electrons of the FL$^\ast$ phase and will be studied now.

% %%%%%%%%%%%%%%%%%%%%%%%%%%%%%%%%%%%%%%%%%%%%%%%%%%%%%%%%%%%%%%%%%%%%%%%
% \begin{figure}[t!]
% \centering
% \includegraphics[width=\columnwidth]{lambda_crit_AFM.png}
% \caption{\todo{1. Pls change aspect ratio, reduce figure height by $\sim$25\%. 2. Update $x$-axis so that half filling is $n=1$.} Phase diagram of the AFM Hubbard--Kitaev model as a function of critical scale $\Lambda_c$ vs.\ filling $n$.
% A magnetically ordered phase (SDW) at and near van-Hove filling $n_{\rm vH}=\red{2\times}0.625$ is surrounded on both sides by a chiral $d$-wave superconductor. Shown are results for $U/t=2$, 2.5 and 3.}
% \label{fig:FRG-AFM}
% \end{figure}
% %%%%%%%%%%%%%%%%%%%%%%%%%%%%%%%%%%%%%%%%%%%%%%%%%%%%%%%%%%%%%%%%%%%%%%%

{\it FRG analysis of effective model.---}
\label{sec:fRG}
We analyze the leading instabilities of the Hubbard--Kitaev model by virtue of the truncated-unity FRG (TUFRG)\,\cite{metzner2012,platt2013} method, using the recently published FRG codebase divERGe\,\cite{diverge}. FRG interpolates between the bare Hubbard--Kitaev interaction and a low-energy two-particle interaction vertex, by means of iteratively integrating a flow equation.
Leading divergences are then extracted from the two-particle interaction vertex by solving the linearized gap equation.
In the employed truncated-unity formulation, the interaction vertices are expressed in terms of the lattice harmonics\,\cite{lichtenstein2017,husemann2009,wang2012,diverge,beyer2022, bunney2024} giving us precise access to the explicit forms of the leading instabilities. We calculated up to possible forth-nearest-neighbor pairings, with a transfer momentum discretization of $108^2$ momentum points.
A further $51^2$ refinement is used to evaluate the Green's functions\,\cite{beyer2022}.

Given the model construction, we are interested in weak coupling, $U\ll W$, with bandwidth $W = 6t$.
In practice, we operate at $1/3 \leq U/W\leq 1/2$ in order to obtain numerically tractable instabilities; for the same reason we focus on band fillings in the vicinity of the van-Hove singularity located $n_{\rm vH}=5/4$ where effects of weak interactions are strongest.

In Fig.\,\ref{fig:FRG} we present the TUFRG phase diagrams for the (a) AFM and (b) FM Hubbard-Kitaev models. We find a spin-density wave (SDW) at and near van-Hove filling, surrounded at lower and higher fillings by superconducting phases.
The results for $U/t=2, 2.5, 3$ are qualitatively similar, Fig.\,\ref{fig:FRG}; stronger interactions slightly increase the filling range of the magnetic phase, but mostly increase the critical scale $\Lambda_c$ and thus stabilize the symmetry-broken states. Importantly, our numerical results are stable over an extended $U$ range and do not require any fine-tuning.

For the FM Hubbard-Kitaev model with $U/t=2$ and fillings $\nc\leq 0.6$, the FRG does not find any instability down to the lowest cutoff considered, $\Lambda_{\text{min}} = 10^{-6}$. A finer momentum discretization would be needed to determine whether an instability occurs at even smaller scales.

The SDW phases in both phase diagrams correspond to commensurate (incommensurate) spin order at (near) van-Hove filling, respectively; further details are relegated to the SM \cite{suppl}.

%%%%%%%%%%%%%%%%%%%%%%%%%%%%%%%%%%%%%%%%%%%%%%%%%%%%%%%%%%%%%%%%%%%%%%%
{\it Mean-field theory of fractionalized superconductivity.---}\label{sec:mft}
In order to study the properties of the emergent SC$^\ast$ state, we treat the Hubbard--Kitaev model using BCS-type mean-field theory.
TUFRG delivers an effective bilinear interaction corresponding to the superconducting instability, which we can directly input into a Bogoliubov--de Gennes (BdG) Hamiltonian.
We define the superconducting order parameter \cite{sigrist1991},
\begin{equation}
    \Delta = (\Psi \, \sigma_0 + \boldsymbol{d} \cdot \boldsymbol{\sigma} ) \; i \sigma_y \, ,
\end{equation}
where the identity matrix $\sigma_0$ and the Pauli matrices $\boldsymbol{\sigma} = (\sigma_x, \sigma_y, \sigma_z)$ act on the spin subspace.
All the spatial (and therefore sublattice) pairing information is contained in the coefficients; $\Psi$ is the spin-singlet pairing and $\boldsymbol{d}=(d_x,d_y,d_z)$ are the three spin-triplet pairings.
The representation theory of the effective model reveals that the corresponding {\it total} group is $O_h \times \Ztwo$; for details see SM.

\begin{figure}[t!]
\centering
\includegraphics{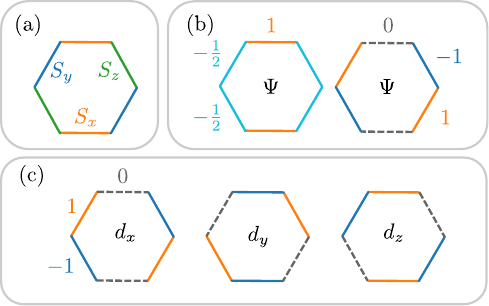}
\caption{
(a) Bond-dependent interactions of the Kitaev model: bonds with ``$S_x$'' correspond to $K S_i^x S_j^x$ etc. (b) Dominant real-space pairings of the two degenerate $d$-wave states found in the AFM Hubbard--Kitaev model ($\Psi$ indicates spin-singlet). Red/blue colors refer to positive/negative pairing amplitude. (c) Same as (b), but for the three degenerate $p$-wave states found in the FM Hubbard--Kitaev model ($d_x, d_y, d_z$ indicate the three spin-triplet components).
}
\label{fig:real-space-pairings}
\end{figure}

\textit{AFM case.}--- The SC state is composed of two-fold degenerate $d$-wave states, which form a set of basis states of the $E_{2g}$ irrep of the group $O_h \times \Ztwo$.
We can write the superconducting term of the mean-field Hamiltonian,
as known from the standard Hubbard model\,\cite{nandkishore2012, black-schaffer2014,bunney2024}, as
\begin{equation}
\label{afm_sc}
    \mathcal{H}_{MF}^{(-)} = \sum_{\langle i,j \rangle_\alpha, \, s s^\prime} \Delta^\alpha_m (i \sigma^\pd_y)_{s s^\prime}^\pd c_{i s}^\dagger c_{j s^\prime}^\dagger + \text{H.c.}\ .
\end{equation}
The two sets of $\vec{\Delta}_m = (\Delta_m^1, \Delta_m^2, \Delta_m^3)$ with $m\in\{d_{x^2-y^2}, d_{xy}\}$ %($m=1,2$)
are given by:
\begin{equation}
    \vec{\Delta}_{d_{x^2 - y^2}} \propto \bigg( 1, -\frac{1}{2}, -\frac{1}{2} \bigg) \; {\rm ~and~} \;
    \vec{\Delta}_{d_{xy}} \propto ( 0, 1, -1) .
\end{equation}
The dominant superconducting pairing is between nearest neighbors, with a relatively small next-nearest neighbor contribution and negligible pairings on third- and forth-nearest neighbors. This pairing corresponds to the chiral $d$-wave state well-known from other hexagonal-lattice systems with $E_{2g}$ irrep\,\cite{honerkamp2003, nandkishore2012, bunney2024}. Its real-space structure is shown in Fig.\,\ref{fig:real-space-pairings}(b), while the momentum-space order parameters of the different spin and orbital states are plotted in the Supplementary Material (SM) \cite{suppl}. The SM also contains the rationale behind the $d$-wave naming convention for the honeycomb lattice.

%%%%%%%%%%%%%%%%%%%%%%%%%%%%%%%%%%%%%%%%%%%%%%%%%%%%%%%%%%%%%%%
%%%%%%%%%%%%%%%%%%%%%%%%%%%%%%%%%%%%%%%%%%%%%%%%%%%%%%%%%%%%%%%

The topological properties of the superconducting states can be revealed by performing a Bogoliubov transformation of $\mathcal{H} = \mathcal{H}_t + \mathcal{H}_{MF}^{(-)}$. To find the ground state, the superposition of the two superconducting states with lowest energy must be found, i.e., the free energy be minimized. For this particular case the chiral $d\pm id$ superposition has lowest energy\,\cite{black-schaffer2014,bunney2024}, $\vec{\Delta}_{d \pm id} = \vec{\Delta}_{d_{x^2 - y^2}} \pm i \vec{\Delta}_{d_{xy}}$.

Through a calculation of the Berry curvature using numerical integration\,\cite{fukui2007}, we find  Chern number $\mathcal{C} = 4$ for both superconducting phases. $\mathcal{C}=4$ matches the results in the literature for a chiral $d$-wave superconductor with dominant nearest-neighbor pairing\,\cite{crepieux2023,bunney2024}. In the SM, we also show examples of corresponding ribbon spectra with four chiral in-gap states per edge, as expected from bulk--boundary correspondence.

\textit{FM case.}---The three-fold degenerate superconducting instability leads to a spin-triplet $p$-wave state primarily on the nearest-neighbor bonds. These three states span the $T_{1u}$ irrep of the total group. The pairing part of the mean-field Hamiltonian can be written as
\begin{equation}
\label{fm_sc}
    \mathcal{H}_{MF}^{(+)} = \sum_{\langle i,j \rangle_\alpha, \, s s^\prime} \Delta_m^\alpha ( \sigma_m^\pd \, i \sigma_y^\pd )_{s s^\prime}^\pd c_{i s}^\dagger c_{j s^\prime}^\dagger + \text{H.c.} \, ,
\end{equation}
where $m \in \{x, y, z\}$ for each of the three degenerate SC states. For instance, when $\vec{\Delta}_z \propto (1, -1, 0)$, the pairing amplitude $\Delta^3_z = 0$ across the Kitaev $z$ bond, Eq.\,\eqref{eff_int}, as depicted in Fig.\,\ref{fig:real-space-pairings} (c). The other two degenerate states are obtained by permuting the entries of $\vec{\Delta}$, fixing $\Delta_m^\alpha = 0$ for $m= \alpha$.
In the same way we classified the AFM instability as $d$-wave above, we can classify this SC order parameter as $p$-wave, see the SM for further details. We emphasize that this $p$-wave state is different from the standard chiral $p+ip$ state with $E_{1u}$ irrep known from the $D_{6h}$ point group.

In order to find the stable superconducting state, we perform a Landau-Ginzburg analysis\,\cite{volovik1985,nandkishore2012} (see SM \cite{suppl} for details) which identifies the
so-called chiral superposition\,\cite{agterberg1999, carlstrom2011, maiti2013}, $\vec \Delta_x + \varepsilon \vec \Delta_y + \varepsilon^* \vec \Delta_z$, with $\varepsilon=e^{i2\pi/3}$ as energetically most favorable.
The resulting state is fully gapped. By performing the analogous topological analysis, we find $\mathcal{C}=4$ in the superconducting phase for fillings {\it below} the SDW phase. For the phase {\it above} the SDW phase, the Chern number is $\mathcal{C}=2$. In the SM, we show examples of corresponding ribbon spectra.

In summary, we find for the AFM (FM) case fractionalized topological superconductivity with spin-singlet (spin-triplet) pairing and Chern number $\mathcal{C}=4$ (Chern numbers $\mathcal{C}=4$ and $\mathcal{C}=2$, respectively).

%%%%%%%%%%%%%%%%%%%%%%%%%%%%%%%%%%%%%%%%%%%%%%%%%%%%%%%%%%%%%%%%%%%%%%%

\paragraph{Transition to conventional superconductivity.---}
\label{sec:trans}

Given the results of Refs.~\onlinecite{seifert2018,choi2018}, the SC$^\ast$ phase established in this paper will transit into a non-fractionalized superconducting phase upon increasing $\JK$. The $T=0$ SC$^\ast$--SC transition is a confinement transition; it can be either of first order or continuous. In the latter case, the simplest theory is that of the condensation of a bosonic field corresponding to Kondo screening \cite{senthil2003,senthil2004}.
Given the $\Ztwo$ nature of the spin liquid's gauge field, this quantum phase transition is of Ising type \cite{chowdhury2013}.
Since the normal-state Fermi surface reconstructs at the parent FL$^{\ast}$--FL transition, the fermiology of the two superconducting states (Chern numbers etc.) can be expected to be different.
Interestingly, the $\Ztwo$ confinement transition is only well defined at $T=0$ in two space dimensions, such that the continuous SC$^\ast$--SC transition becomes a crossover between two quasi-long-range-ordered states at finite $T$.
A detailed study of this transition is left for future work; some of its signatures have been considered in Ref.~\onlinecite{chowdhury2013}.

%%%%%%%%%%%%%%%%%%%%%%%%%%%%%%%%%%%%%%%%%%%%%%%%%%%%%%%%%%%%%%%%%%%%%%%

{\it Summary.---}
We have shown that the Kitaev--Kondo lattice admits a fractionalized superconducting phase SC$^\ast$, where BCS-type superconductivity of conduction electrons coexists with a spin-liquid background and an emergent gauge structure.
This phase is separated from a more conventional (heavy-fermion) superconductor by a confinement transition driven by Kondo screening (or its breakdown); this transition is hence the superconducting version of the FL$^\ast$--FL transition discussed in the context of Kondo-breakdown quantum criticality.
The pairing of SC$^\ast$ is driven by the magnetic excitations of the Kitaev spin liquid, i.e., by Majorana glue. Our FRG analysis has revealed the pairing symmetry to be topologically non-trivial and of chiral spin-singlet (chiral spin-triplet) type for AFM (FM) Kitaev interactions, for band fillings away from but near van-Hove filling.
Our arguments for an SC$^\ast$ phase straightforwardly extend to other Kitaev spin liquids; in three-dimensional systems \cite{hermanns2016} where true superconductivity exists at finite $T$ the $\Ztwo$ confinement transition can be expected to persist at finite $T$ as well.

More broadly, we have established a new pairing mechanism -- the fluctuations of a fractionalized spin liquid -- leading to topological superconductivity. Future work should establish such pairing beyond the perturbative limit and in non-Kitaev systems. Experimentally, studying hetero-systems comprised of frustrated local moments and conduction electrons appears most promising to identify the phases proposed here.

\textit{Note added.}---
While completing this work, we became aware of Ref.~\onlinecite{lundemo2024} which also studies weak-coupling superconductivity in the FM Kitaev--Kondo lattice. Their results are broadly consistent with ours, however, they only consider small dopings near half filling, related to the phase we find at fillings below the van-Hove singularity, with Chern number $2$ instead of 4.
A very different type of fractionalized superconductor appears in Ref.~\onlinecite{panigrahi2024}: This paper 
investigates a Kondo lattice based on the Yao-Lee model where the gauge field is deconfined across the entire phase diagram by construction.

%%%%%%%%%%%%%%%%%%%%%%%%%%%%%%%%%%%%%%%%%%%%%%%%%%%%%%%%%%%%%%%%%%%%%%%
\acknowledgments

%We acknowledge instructive discussion with ... M. Bunney
We would like to thank L. Klebl and J. Profe for their help with the use of their FRG codebase\,\cite{diverge} and fruitful physics discussions, as well as J. Knolle for providing the data of Ref.~\onlinecite{knolle2014}. M.B. would like to thank A. Savvinos for her insights around the group theoretic aspects of this project.
This work was funded by the Deutsche Forschungsgemeinschaft (DFG) through SFB 1143 (project id 247310070), SFB 1238 (project id 277146847), the W\"urzburg-Dresden Cluster of Excellence on Complexity and Topology in Quantum Matter -- \textit{ct.qmat} (EXC 2147, project id 390858490) and the Emmy Noether Program (SE 3196/2-1, project id 544397233). S.R.\ acknowledges support from the Australian Research Council through Grant No.\ DP200101118 and DP240100168. The authors gratefully acknowledge the scientific support and HPC resources provided by the Erlangen National High Performance Computing Center (NHR@FAU) of the Friedrich-Alexander-Universität Erlangen-N\"urnberg (FAU) under the NHR project ``k102de--FRG". NHR funding is provided by federal and Bavarian state authorities. NHR@FAU hardware is partially funded by the DFG – 440719683.

%%%%%%%%%%%%%%%%%%%%%%%%%%%%%%%%%%%%%%%%%%%%%%%%%%%%%%%%%%%%%%%%%%%%%%%
%%%%%%%%%%%%%%%%%%%%%%%%%%%%%%%%%%%%%%%%%%%%%%%%%%%%%%%%%%%%%%%%%%%%%%%
%%%%%%%%%%%%%%%%%%%%%%%%%%%%%%%%%%%%%%%%%%%%%%%%%%%%%%%%%%%%%%%%%%%%%%%

\bibliography{kitaev_bib}

\begin{thebibliography}{58}
\bibitem [{foo()}]{footnote}%
  \BibitemOpen
  \href@noop {} {}\bibinfo {note} {Note that for the alternative definition of
  the spin operators, $s_i^\alpha = c_{i\sigma}^\dag
  \tau_{\sigma\sigma'}^\alpha c_{i\sigma'}^{\phantom{\dag}}$, $\kappa$ is
  consequently reduced by a $1/4$, i.e., $\kappa =\mp 0.14828$.}\BibitemShut
  {Stop}%
\end{thebibliography}

\end{document}

\title{
Supplementary Material:\texorpdfstring{\\}{}
Fractionalized superconductivity from Majorana glue in the Kitaev--Kondo lattice
}

\author{Matthew Bunney}
\affiliation{School of Physics, University of Melbourne, Parkville, VIC 3010, Australia}
\affiliation{Institute for Theoretical Solid State Physics, RWTH Aachen University, 52062 Aachen, Germany}
\author{Urban F.\ P.\ Seifert}
\affiliation{Institut f\"ur Theoretische Physik, Universit\"at zu K\"oln, Z\"ulpicher Str. 77a, 50937 K\"oln, Germany}
\author{Stephan Rachel}
\affiliation{School of Physics, University of Melbourne, Parkville, VIC 3010, Australia}
\author{Matthias Vojta}
\affiliation{Institut f\"ur Theoretische Physik and W\"urzburg-Dresden Cluster of Excellence ct.qmat, Technische Universit\"at Dresden, 01062 Dresden, Germany}

%%%%%%%%%%%%%%%%%%%%%%%%%%%%%%%%%%%%%%%%%%%%%%%%%%%%%%%%%%%%%%%%%%%%%%%

\date{\today}

\maketitle
%%%%%%%%%%%%%%%%%%%%%%%%%%%%%%%%%%%%%%%%%%%%%%%%%%%%%%%%%%%%%%%%%%%%%%%

\tableofcontents

\section{Magnetic order}

Previous FRG works have identified the leading divergence of the standard honeycomb-lattice Hubbard model at van-Hove filling ($\nc=5/4$) as having three peaks in the transfer momentum spectrum, at the three inequivalent $\boldsymbol{M}$ points\,\cite{kiesel2012, wang2012, wu2013}. Subsequent TUFRG work showed that this result is stable over a larger phase space, including some small nearest-neighbor and next-nearest-neighbor density--density interactions $V$\,\cite{delapena2017}.

Symmetry under spatial rotation requires that all three peaks be equally divergent, and $\SUtwo$ spin symmetry requires that we have an additional threefold spin degeneracy. 
%In other words, each $\boldsymbol q = \boldsymbol M$ peak has a divergent eigenvalue $\lambda$, where the leading divergent eigenvalue is degenerate across the three peaks, as well as three-fold degenerate within each peak.
Hence, solving the linearized gap equation yields nine degenerate solutions
\begin{equation}
\label{eqn:sdw}
    \mathcal{H}_{\text{SDW}} = \sum_{\boldsymbol k s s^\prime} c_{\boldsymbol k + \boldsymbol M, \, s}^\dagger \; \sigma^\alpha_{ss^\prime} \; c_{\boldsymbol k, \, s^\prime}^\pd
\end{equation}
for the three degenerate $\boldsymbol M$ points and the three Pauli spin matrices $\sigma^\alpha$. Further study into this model using a peturbative expansion of the Landau-Ginzburg free energy functional of the mean-field model found that the ground state was a collinear order with an enlarged eight site unit-cell and non-uniform spin moments, forming a half-metal\,\cite{nandkishore2012_SDW}.

Deviating from van-Hove filling destroys the perfect nesting of the Fermi surface. For fillings above van-Hove filling, $\nc > 5/4$, the peak splits and moves in $\boldsymbol q$ space along the line $\boldsymbol{M}\boldsymbol{K}$. For fillings below the van-Hove singularity, $\nc < 5/4$, the splitting happens along the line $\boldsymbol{M} \boldsymbol{\Gamma}$ instead. 

\begin{figure}[b!]
    \centering
    \includegraphics[width=\columnwidth]{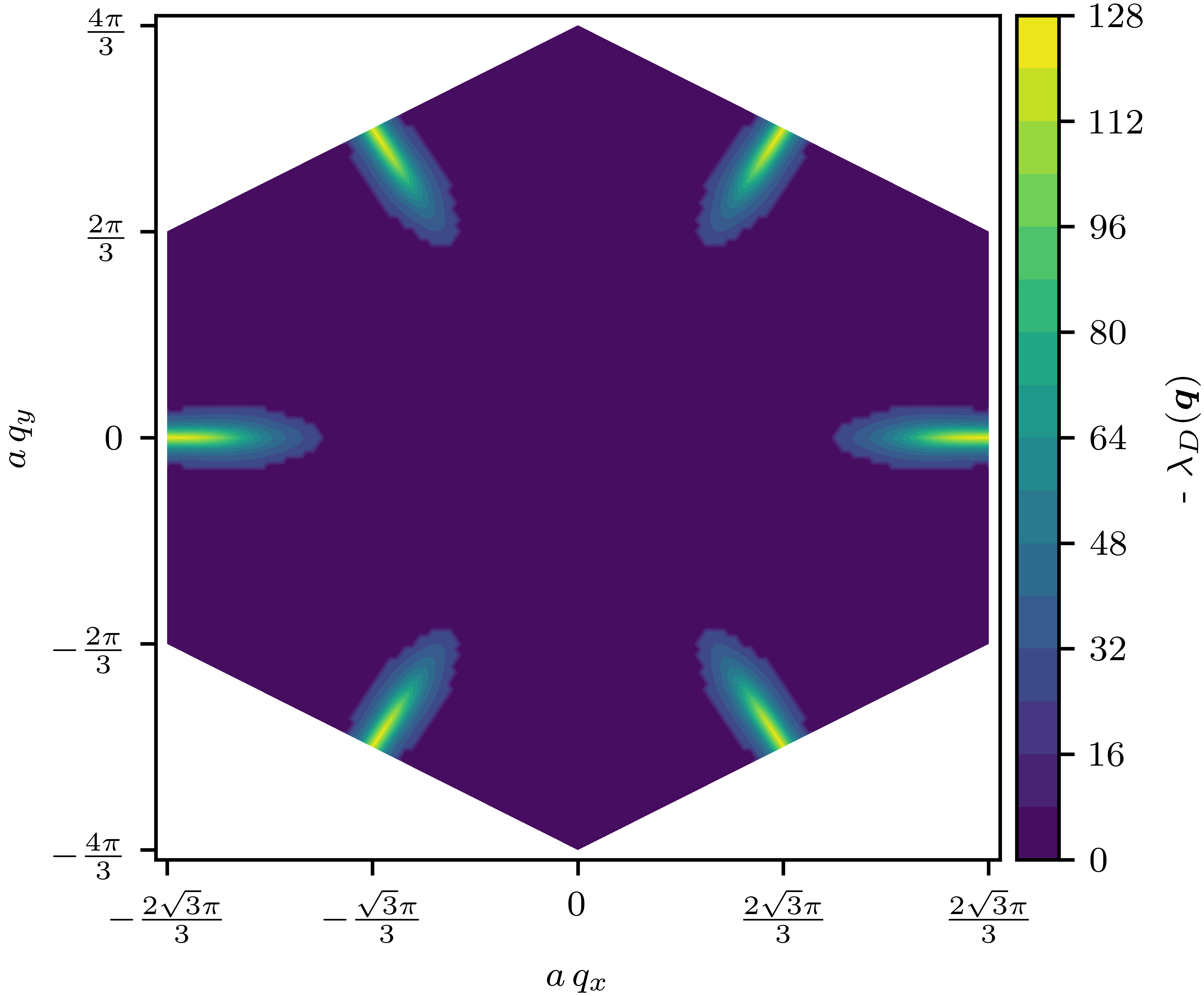}
    \caption{Leading negative particle--hole eigenvalue for each transfer momentum $\mathbf{q}$, corresponding to the magnetic instability (SDW). Parameters are $U=3.0$ and $\nc=5/4$, \textit{i.e.,} at van-Hove filling. The most negative eigenvalues are at the three $\mathbf{M}$ points.}
    \label{fig:mag_evals}
\end{figure}

Using the effective interaction given in Eq.\,(6) %\ref{eff_int} 
as the bare interaction in the FRG flow breaks the $\SUtwo$ spin rotation, thereby reducing the degeneracy of the leading divergence. 
For the AFM Hubbard--Kitaev model, the peak structure is similar to the $\SUtwo$-symmetric density--density case, \textit{i.e.,} there are three peaks located at the three inequivalent $\boldsymbol q = \boldsymbol M$ points, which have degenerate leading eigenvalues. The leading attractive eigenvalue of the linearized gap equation is plotted in Fig.\,\ref{fig:mag_evals}, which is identical for the AFM Hubbard--Kitaev model and the standard $\SUtwo$-symmetric Hubbard model. The $\boldsymbol q$ space structure of the incommensurate SDWs away from van-Hove filling is also the same for the two cases.
The difference to the standard Hubbard model is that the leading eigenvalue is only six-fold degenerate. The corresponding leading instabilities take the form given in Eq.\,\eqref{eqn:sdw}, where the index of the Pauli matrix $\alpha$ does not equal the $\alpha$ of the spin interaction on the bond that is collinear to the vector $\boldsymbol M$. Explicitly, if we consider the Kitaev interaction to be $\propto S^x S^x$ along the bond in the $\hat{\boldsymbol{x}}$ direction, as per Fig.\,2 %\ref{fig:real-space-pairings}
(a), then in the transfer momentum subspace $\boldsymbol q = \boldsymbol M = M \hat{\boldsymbol x}$ the solutions to the linearized gap equation Eq.\,\eqref{eqn:sdw} are indexed by $\alpha = y$ and $z$. The same logic for the other two $\boldsymbol M$ momenta subspaces gives a divergence for a six-fold degenerate solution.
Applying the analysis in  Ref.~\onlinecite{nandkishore2012_SDW} to this SDW instability, one sees that the AFM Kitaev interaction does not qualitatively modify the SDW state found for the standard Hubbard model.

For the FM Hubbard--Kitaev model, the situation is reversed. There is now a three-fold degenerate leading instability, and the index of the Pauli matrix $\alpha$ is equal to the $\alpha$ of the spin interaction on the bond that is collinear to the vector $\boldsymbol M$.
%
% In conclusion, it might still be possible that the chiral SDW \cite{li2012,black-schaffer2014} is the correct magnetic ground state of the AFM/FM Hubbard-Kitaev model around van-Hove filling. However, a detailed analysis to answer this question convincingly is beyond the scope of this work.
Unlike the AFM Hubbard--Kitaev SDW instability, the SDW instability of FM Hubbard--Kitaev model cannot form a collinear spin state, as the SDW OP at different $\boldsymbol M$ cannot align with the same spin pairing $\sigma^\alpha$. Instead, it may be possible that a chiral spin density wave is the ground state of the system\,\cite{li2012, black-schaffer2014}; this issue is beyond the scope of the present paper.

\section{Symmetry of superconducting order parameters}

\begin{figure}[t]
    \centering
    \includegraphics[width=\columnwidth]{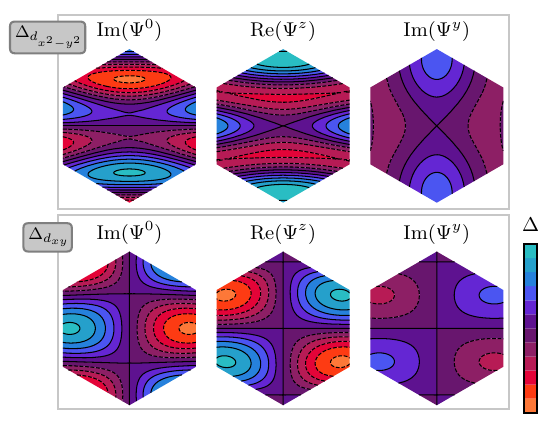}
    \caption{
        Momentum-space structure of the superconducting OP as obtained from FRG for AFM Kitaev interaction. Parameters are $U=3.0$, $\nc=0.590$ where $\Delta_{d_{x^2-y^2}}$ and $\Delta_{d_{xy}}$ are degenerate.
        Only the non-zero components are shown. A $d$-wave structure can be observed in the orbital-swap-symmetric pairings $\Psi^z, \Psi^y$.
        %The OP can be classified as $d$-wave by consider the symmetry around the $\Gamma$ point (which is at the center of the plotted hexagonal Brillouin Zones) of orbital pairings which are point-group symmetric, which in this case are the right two pairings of each of the degenerate states. The relative magnitudes of the OPs indicates that the (inter-sublattice) nearest-neighbor pairings (the left two pairings) have a larger amplitude than the (intra-sublattice) next nearest-neighbor pairings (pairing on the right).
        }
    \label{fig:afm_sc}
\end{figure}

\begin{figure}[t]
    \centering
    \includegraphics[width=\columnwidth]{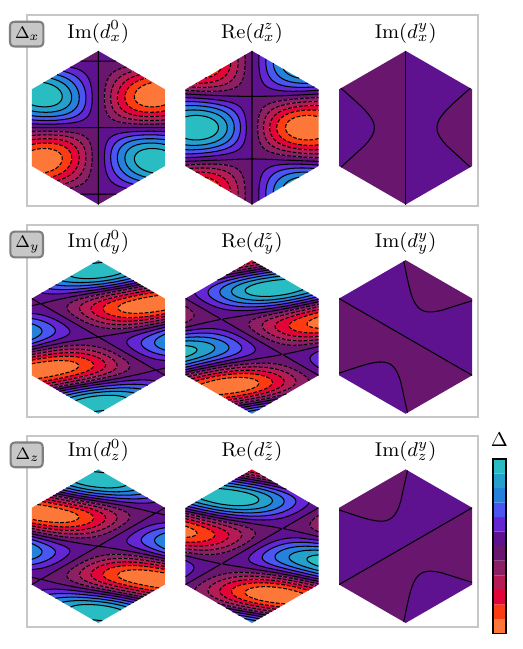}
    \caption{
        As in Fig.~\ref{fig:afm_sc}, but now for FM Kitaev interaction where $\Delta_x$, $\Delta_y$ and $\Delta_z$ are degenerate. A $p$-wave structure can be observed in the orbital-swap-symmetric pairings $d_i^z, d_i^y$.
        %The OP can be classified as $p$-wave by consider the symmetry around the $\Gamma$ point (which is at the center of the plotted hexagonal Brillouin Zones) of orbital pairings which are point-group symmetric, which in this case are the right two pairings of each of the degenerate states. The relative magnitudes of the OPs indicates that the (inter-sublattice) nearest-neighbor pairings (the left two pairings) have a larger amplitude than the (intra-sublattice) next nearest-neighbor pairings (pairing on the right).
        }
    \label{fig:fm_sc}
\end{figure}

It is conventional to use the symmetry of the momentum-space pairing to characterize the superconducting OP. This is a straightforward task when there is only one orbital/sublattice per unit cell, where the OP can be written in the form
\begin{equation}
    \hat \Delta (\boldsymbol k) = ( \Psi ( \boldsymbol k ) \hat{\sigma}_0 + \boldsymbol{d} ( \boldsymbol{k} ) \cdot  \hat{\boldsymbol{\sigma}} ) i \hat{\sigma}_y\ .
\end{equation}
The hat on the OP and the Pauli matrices $\hat{\sigma}_0, \hat{\boldsymbol{\sigma}} = (\hat{\sigma}_x, \hat{\sigma}_y, \hat{\sigma}_z)$ ($\hat{\sigma}_0$ is the identity) indicate that it is a matrix over spin indices of the Cooper pair, \textit{i.e.}, $\Delta^{\phantom{\dagger}}_{s s'} (\boldsymbol{k}) c^\dagger_{\boldsymbol{k} s} c^\dagger_{-\boldsymbol{k} s'}$. The crystal point-group symmetry of the superconducting OP then restricts the possible forms of the functions $\Psi (\boldsymbol{k}), \boldsymbol{d} (\boldsymbol{k}) = (d_x (\boldsymbol{k}), d_y (\boldsymbol{k}), d_z (\boldsymbol{k}))$ to one of the group's irreducible representations. It is then generally possible to classify these functions by their angular symmetry around the $\Gamma$ point with $\Psi ( \alpha \boldsymbol{k}) = \alpha^\ell \Psi ( \boldsymbol{k} )$, with $\ell = 0, 1, 2, ...$ corresponding to $s, p, d, ...$ pairing (where we swap $\Psi$ for $d_i$ in the case of a spin-triplet pairing)\,\cite{sigrist1991}.

When there is more than one site per unit cell, we must proceed with caution. The Cooper pairs now carry a sublattice index $o = a, b$ for each of the two sublattices on the honeycomb lattice. This affords the OP an extra level of structure: a structure over the sublattice index. We can breakdown each of the spin components of the order parameter $\Psi, \boldsymbol{d}$ listed above further into their respective orbital components analogously to how the spin structure was:
\begin{eqnarray}
    \hat{\Psi} = (\Psi^0 \hat{\tau}_0 + \Psi^x \hat{\tau}_x + \Psi^y \hat{\tau}_y + \Psi^z \hat{\tau}_z ) i \hat{\tau}_y \\[5pt]
    \hat{d}_i = (d_i^0 \hat{\tau}_0 + d_i^x \hat{\tau}_x + d_i^y \hat{\tau}_y + d_i^z \hat{\tau}_z ) i \hat{\tau}_y
\end{eqnarray}
where the hats on the spin components $\hat{\Psi}, \hat{d}_i$ and the Pauli matrices $\hat{\tau}_0, \hat{\sigma}_x, \hat{\tau}_y, \hat{\tau}_z$ now indicate that they act on \textit{orbital} or \textit{sublattice} space. Note that we distinguish the spin index on our components as the lower index, and the orbital as the upper index. Explicitly, we can write these orbital pairings out in terms of the Cooper pairs, for example the orbital antisymmetric pairing $\Delta^0_{s s'} ( a^\dagger_{ \boldsymbol k s} b^\dagger_{ -\boldsymbol k s'} - b^\dagger_{ \boldsymbol k s} a^\dagger_{ -\boldsymbol k s'} )$.

Superconducting OPs with additional sublattice structure can then be classified according to the angular momentum around the $\Gamma$ point of each components, as above, as long as the sublattice pairing of the component is trivial under all crystal point-group symmetries. In the case of the honeycomb lattice, this is true for two of orbital pairings $\Psi^i / \boldsymbol{d}^i$ with $i = y, z$, but not for the pairings $i = 0, x$, as these pairings are odd under group symmetries which swap the two sublattices.

Figs.\,\ref{fig:afm_sc} and \ref{fig:fm_sc} show the two SC OPs discussed in the main text. In both figures, the OP is classified as $d$-wave and $p$-wave, respectively, as shown by the orbital $y$ and $z$ components. Note that the chosen breakdown of the orbital structure of the OP allows us to identify the $x$ and $y$ components as intra-orbital pairings, and $0$ and $z$ as inter-orbital pairings. The outcome of the renormalization-group flow is then that the inter-orbital pairing on nearest-neighbor atoms dominates the overall SC pairing in both cases.

\section{Symmetries of the Hubbard--Kitaev model and superconducting basis states}

The standard Hubbard model on the honeycomb lattice has a total symmetry group $D_{6h} \times \SUtwo$, combining the point and spin rotation groups. The aim of this section is to discuss the reduction of symmetry due to introducing the Kitaev interaction term\,(6). We will identify the group structure, which is a necessary step in identifying possible superconducting instabilities.

When a model exhibits an $\SUtwo$ spin-rotation symmetry, it remains invariant under rotation of the electrons $c_{i r} \rightarrow U_{rs} (\theta, \hat{\boldsymbol n}) c_{i s}$. Here $U_{rs} (\theta, \hat{\boldsymbol n}) = ( \exp [ i \theta ( \hat{\boldsymbol n} \cdot \boldsymbol \tau)/2])_{rs}$ is the usual form of an $\SUtwo$ rotation of a spin 1/2 particle by angle $\theta$ around normal vector $\hat{\boldsymbol n}$, with $\boldsymbol \tau = (\tau^x, \tau^y, \tau^z)$ as the 3-vector of Pauli matrices. For a bilinear $c_{i r} \tau^\alpha_{rs} c_{j s}$, $\SUtwo$ rotation is equivalent to a rotation of the spin quantization axes, \textit{i.e.}, mapping $\tau^\alpha \rightarrow R_{\alpha \beta} (\theta, \hat{\boldsymbol n}) \, \tau^\beta$, where $R_{\alpha \beta} (\theta, \hat{\boldsymbol n})$ is a three-dimensional rotation acting on three-vector $\boldsymbol \tau$.

The Kitaev term only allows spin rotations that leave all quadratic products of the Pauli matrices $\tau^\alpha \tau^\alpha$ invariant. These are the three $\pi$ rotations each about the quantization axes $\hat{U} (\pi, \hat{\boldsymbol{n}})$ for $n = x,y,z$. This reduces the spin rotation symmetry from $\SUtwo$ to the Klein 4-group $K_4 \simeq \Ztwo \times \Ztwo$ \cite{baskaran08,you2012,liu24}.

The point-group symmetries are also altered, as spatial rotations must also be accompained by spin rotations to remain symmetries of the model. In particular, we can define for the generators of $D_{6h}$\,\cite{seifert2018}:
\begin{itemize}
    \item $C_6$ rotations must be accompanied by a permutation of the spin quantization axes $(x,y,z) \rightarrow (y,z,x)$. This can be thought of as the spin rotation $\hat{U} ( 2 \pi/3, (1,1,1)/\sqrt{3} )$.
    \item Mirror planes must swap two of the spin axes, \textit{e.g.} $\sigma_x$ acts $(-x, -z, -y)$. This is the spin rotation $\hat{U} ( \pi, (0,1,-1)/\sqrt{2})$.
    \item Inversion $i$, which does not require a spin rotation.
\end{itemize}
The choices of the accompanying spin rotations are unique up to the $K_4$ spin rotations above, or in other words, the choice of spin symmetries are unique up to the quotient of the Klein 4-group. What is important, however, is that these choices of spin rotations ensure that $D_{6h}$ remains a group (closed under multiplication).

This effective two-particle spin-orbit coupling means that the structure of the group is $D_{6h} \ltimes K_4$ -- \textit{i.e.}, a semi-direct rather than a direct product. The important consequence of this structure is that irreps of the larger group cannot be understood as a product of irreps of the two factors\,\cite{serre1977}. It is then useful to identify the group structure, \textit{i.e.}, what group is the semi-direct product isomorphic to, which was found to be $O_h \times \Ztwo$.

While this group has 20 conjugacy classes (and therefore 20 irreps), we can understand the character table by considering the subgroup $\mathcal{S}_4$ (the symmetric group). This is because the whole group $O_h \times \Ztwo \simeq \mathcal{S}_4 \times \Ztwo \times \Ztwo$, so the characters and irreps of the larger group can be derived trivially from those of $\mathcal{S}_4 \simeq C_{3v} \ltimes K_4$, which is generated by combined spatial/spin symmetries $C_3$, $\sigma_x$ and the spin-only symmetries $a = \hat{U} (\pi, \hat{\boldsymbol{x}}), \; b = \hat{U} (\pi, \hat{\boldsymbol{y}})$. The character table is given in Tab.\,\ref{tab:s4_character_table}. The first column indicates the irrep, and the following five are the five conjugacy classes on the group. What is notable is that the spin symmetries $K_4$ reduce to the one conjugacy class, and we can categorize irreps as either transforming trivially or non-trivially under spin symmetries based on if the character is the same as the identity. The top three irreps are all identical under these spin transformations, and the characters follow those for the irreps of $C_{3v}$. We can see that if we expand out to the full group $O_h \times \Ztwo \simeq D_{6h} \ltimes K_4$, that these irreps have the same character as those of $D_{6h}$, as long as they transform trivially under the $K_4$ spin transformations, which would only be the spin-singlet superconducting pairings (\textit{i.e.}, the superconducting pairing that is antisymmetric upon exchanging the spins). This leaves only the three-dimensional irreps of $O_h \times \Ztwo$ as possible basis functions of the spin triplet (spin exchange symmetric) superconducting states.

\begin{table}[t]
    \centering
    \begin{tabular}{cccccc} \toprule
        {~$\mu$~} & ~$[E]$~ & ~$8 [ C_3 ]~$ & ~$6 [ \sigma_x ]$~ & ~$3 [ a ]$~ & ~$6 [ \sigma_x C_3 a ]$~ \\ \midrule
        $A_1$  & 1 & 1 & 1 & 1 & 1 \\
        $A_2$  & 1 & 1 & $-1$ & 1 & $-1$ \\
        $E$    & 2 & $-1$ & 0 & 2 & 0 \\ \midrule
        $T_1$  & 3 & 0 & 1 & $-1$ & $-1$ \\
        $T_2$  & 3 & 0 & $-1$ & $-1$ & 1 \\ \bottomrule
    \end{tabular}
    \caption{Character table of $C_{3v} \ltimes D_2 \simeq \mathcal{S}_4$.}
    \label{tab:s4_character_table}
\end{table}

We can construct the superconducting basis functions by forming projection operators for each of the irreps\,\cite{bunney2024}. The representation we choose is the Hilbert space of Cooper pairs across nearest neighbors $ \sim e^{i \boldsymbol k \cdot \boldsymbol r} c_{- \boldsymbol k o \sigma} c_{\boldsymbol k o' \sigma'}$, where $\boldsymbol r$ is a nearest-neighbor bond vector, $o, \, o'$ are orbital indices and $\sigma, \sigma'$ are spin indices. The Hilbert space has a total dimension of 24, for 6 unique nearest-neighbor bonds times 4 spin combinations. The basis functions are then the eigenvectors of the projection operators with an eigenvalue of 1. The results are summarized in Tab.\,\ref{tab:basis_functions}. Each row shows a set of basis functions, labelled by irrep of $O_h \times \Ztwo$. The mean-field superconducting Hamiltonian corresponding to each basis function in the table can be written down in the form
\begin{equation}
    \mathcal{H} = \sum_{\boldsymbol k \, oo' \, ss'} \Delta^n e^{i \boldsymbol k \cdot r_n} c_{-\boldsymbol k o s} ( i \sigma^\alpha \sigma^y )^{\phantom{\dagger}}_{ss'} c_{\boldsymbol k o' s'} + \text{H.c.}
\end{equation}
where $\Delta^n$ is the superconducting pairing across the nearest-neighbor bond $r$ pointing from sublattice $o$ to $o'$, and the spin pairing $ i \sigma^\alpha \sigma^y $ is given by $\alpha = 0$ for spin-singlet, and $\alpha=x,y,z$ for spin-triplet pairings $d_x$, $d_y$ and $d_z$, respectively.

\begin{table}[t]
    \centering
    \begin{tabular}{cc} \toprule
        {$\mu$} & Basis Function \\ \midrule
        $A_{1g}$  & \begin{minipage}{0.75\columnwidth}\includegraphics{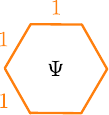} \end{minipage}~\vspace{5pt}  \\ \midrule
        $E_{2g}$  & \begin{minipage}{0.75\columnwidth}\includegraphics{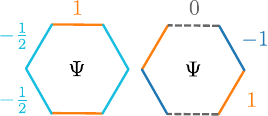} \end{minipage}~\vspace{5pt} \\ \midrule
        $T_{1u}$  & \begin{minipage}{0.75\columnwidth}\includegraphics{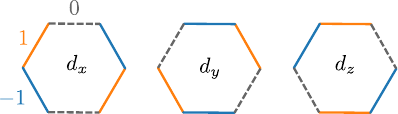} \end{minipage}~\vspace{5pt} \\ \midrule
        \multirow{2}{*}[-28pt]{$T_{2u}$}  & \hspace{0.5cm} \begin{minipage}{0.75\columnwidth}\includegraphics{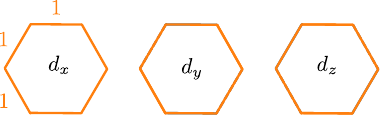} \end{minipage}~\vspace{5pt} \\ \cmidrule{2-2}
                  & \begin{minipage}{0.75\columnwidth}\includegraphics{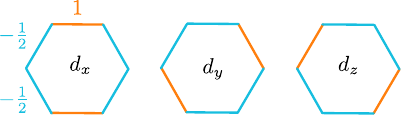} \end{minipage}~\vspace{5pt} \\ \bottomrule
    \end{tabular}
    \caption{Possible superconducting instabilities on nearest-neighbor bonds, organized by irreps of $O_h \times \Ztwo$ (rows). There are two sets of basis functions for the $T_{2u}$ irrep. Red/blue colors refer to positive/negative pairing amplitude.}
    \label{tab:basis_functions}
\end{table}

\section{Landau--Ginzburg analysis}

The divergence of the TUFRG flow at critical scale $\Lambda_c$ indicates a symmetry-breaking phase transition\,\cite{metzner2012}. When this is a superconducting phase transition, we can diagonalize the effective vertex to extract the diverging eigenvector to get the leading superconducting instability, which will transform under an irrep of the symmetry group of the model. We can write this explicitly as
\begin{equation} \label{eqn:sc_instab}
    \hat \Delta = \sum_i \eta_i \hat{\Phi}^i
\end{equation}
where $\eta \in \mathbb{C}$, \textit{i.e.}, we have some complex superposition of the (normalized) superconducting basis functions $\Phi$ of the irrep. The hat indicates that the basis function contains the spin, orbital and spatial structure of the instabilities.
In cases of multiple (degenerate) instabilities, determining the actual ordered state requires a Landau--Ginzburg analysis.

Using the superconducting instability\,\eqref{eqn:sc_instab} as the basis for a Hubbard-Stratonovich transformation, we can transform the effective action that comes out of the FRG flow and integrate out the fermionic degrees of freedom from the partition function to get a Landau--Ginzburg free-energy functional as\,\cite{nandkishore2012, altland2010}:
\begin{equation} \label{eqn:lg-freeen}
    F[\eta, \bar \eta] = - \sum \hat{\Delta} \hat{V}^{-1}_\Lambda \hat{\Delta} + \text{Tr} \; \text{ln} \; \hat{\mathcal{G}}^{-1}
\end{equation}
where $\hat{V}^{-1}$ is the inverse of the effective vertex near the phase transition, and $\mathcal{G}^{-1} = i \omega - H_{\rm BdG}$ is the Gor'kov Green's function, and we trace over all quantum indices. Near the phase transition, we expect the mean-field amplitudes to be small, so we can expand the free-energy functional in orders of the order parameter fields.

For a superconducting instability that spans a three-dimensional irrep of $O_h$ (with the same logic applying to $O_h \times \Ztwo$), such as the $T_{1u}$ found for the ferromagnetic Hubbard--Kitaev model in the main text, the form of the Landau--Ginzburg expansion is restricted by symmetry\,\cite{volovik1985, mineev1999}:
\begin{equation}
    F[\eta, \bar \eta] \approx - \alpha \sum_i | \eta_i |^2 + \beta_1 (\boldsymbol{\eta} \bar{\boldsymbol{\eta}})^2 + \beta_2 | \boldsymbol{\eta}^2 |^2 + \beta_3 \sum_i | \eta_i |^4
\end{equation}
where the coefficients $\alpha$, $\beta_i$ are real. $\alpha$ controls the superconducting phase transition, determining, e.g., the overall superconducting amplitude $|\Delta| = \sum_i | \eta_i |^2$, but does not distinguish relative phases and amplitudes of the individual basis functions $\eta_i$. Instead, this information is determined by quartic terms in the expansion.
The $\beta_i$ coefficients can be calculated from expansion of the Gor'kov Green's function term in Eq.\,\eqref{eqn:lg-freeen} as
\begin{align}
    \beta_1 &= - \text{Tr} \; ( \hat{G}_p \, \hat{\Phi}_i \, \hat{G}_h \, \hat{\Phi}^\dagger_{i} \, \hat{G}_p \, \hat{\Phi}_{j} \, \hat{G}_h \, \hat{\Phi}^\dagger_{j})\ , \\
    \beta_2 &= - \text{Tr} \; ( \hat{G}_p \, \hat{\Phi}_i \, \hat{G}_h \, \hat{\Phi}^\dagger_{j} \, \hat{G}_p \, \hat{\Phi}_{i} \, \hat{G}_h \, \hat{\Phi}^\dagger_{j})\ , \\
    \beta_3 &= - \text{Tr} \; ( \hat{G}^{\phantom{\dagger}}_p \, \hat{\Phi}^{\phantom{\dagger}}_i \, \hat{G}^{\phantom{\dagger}}_h \, \hat{\Phi}^\dagger_{i} \, \hat{G}^{\phantom{\dagger}}_p \, \hat{\Phi}^{\phantom{\dagger}}_i \, \hat{G}^{\phantom{\dagger}}_h \, \hat{\Phi}^\dagger_i)\ .
\end{align}
Here $G_{p/h}$ is the particle/hole Green's function at the critical scale $\Lambda_c$, where we define the Green's functions as used in the TUFRG flow as the corresponding bare Green's function regulated by a sharp cutoff/Heaviside function\,\cite{beyer2022}: $\hat{G}_{p/h} = \hat{G}^0_{p/h} \theta ( |\omega| - \Lambda )$, where $\hat{G}^0_{p} = ( - i \omega - \hat{h} (\boldsymbol k) )^{-1}$ and $\hat{G}^0_{h} = ( - i \omega + \hat{h}^T ( - \boldsymbol k) )^{-1}$ for the Bloch Hamiltonian $\hat{h}$. In this form, we can calculate the $\beta_i$ coefficients numerically -- allowing us to use the full superconducting instabilities that span several pairing distances -- and find that for the $T_{1u}$ superconducting instability obtained in the ferromagnetic Hubbard--Kitaev model, all $\beta_i > 0$, which is enough to determine that the ground state is given (up to permutations and an overall phase factor) by $(\eta_1, \eta_2, \eta_3) = (1, \varepsilon, \varepsilon^*), \, \varepsilon = e^{2 i \pi/3}$\,\cite{volovik1985}, \textit{i.e.} the superconducting ground state is the chiral ground state discussed in the main text, $\hat{\Delta} = \hat{\Phi}_1 + \varepsilon \hat{\Phi}_2 + \varepsilon^* \hat{\Phi}_3$.

\section{Chern number dependence on pairing amplitude}

\begin{figure}[t]
    \centering
    \includegraphics{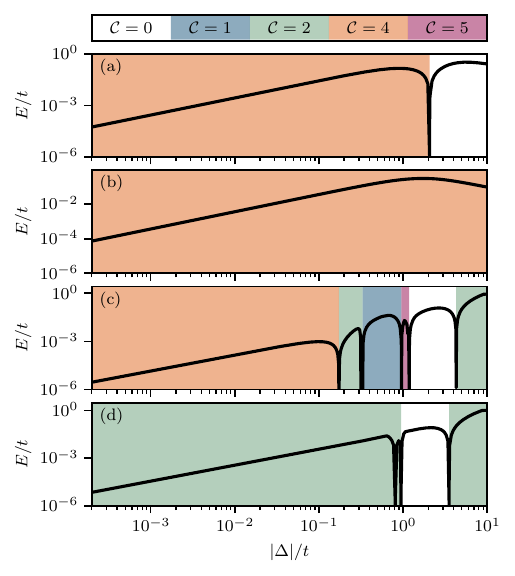}
    \caption{Single-particle gap $E$ of BdG SC as a function of SC amplitude $|\Delta|$. The color encodes the Chern number, as shown in the legend. The chosen parameters in the four panels correspond to those used in Fig.\,\ref{fig:ribbons} for the ribbon spectra. (a) AFM case, $\nc=1.18$. (b) AFM case,  $\nc=1.30$. (c) FM case, $\nc=1.18$. (b) FM case,  $\nc=1.30$. In all plots $U/t=3$.
    % The BdG matrix is constructed from the chiral superposition of the degenerate states, as discussed in the body \todo{reference equations?}. The BdG gap is then found through a numerical minimization algorithm over the Brillouin Zone. The Chern number is numerically calculated for each phase through the Fukui-Hatsugai method, using the BdG eigenvectors.\,\cite{fukui2007}
    }
    \label{fig:bdg_gap}
\end{figure}

When a material undergoes a continuous phase transition into a superconducting phase, the amplitude of the order parameter becomes finite. BCS theory predicts that the OP amplitude at zero temperature is of the order of magnitude of the transition temperature towards the superconducting phase\,\cite{degennes2018, sigrist1991}. Within FRG, the latter can be identified with the critical scale at which the vertex diverges\,\cite{platt2013, bunney2024}.
This consideration is important when forming the BdG matrix, and particularly when it comes to calculating topological properties of the system. This is because the topological band character can change as function of OP amplitude for systems with non-trivial sublattice structure.

This is demonstrated in Fig.\,\ref{fig:bdg_gap} for the superconducting order parameters discussed in the main text. In each of these plots, the BdG matrix is constructed from the normal Bloch Hamiltonian -- with hopping $\mathcal{H}_t$ as in (1)%\eqref{ht}
, and the chemical potential -- and one of the mean-field superconducting Hamiltonians $\mathcal{H}_{\text{MF}}$, either (8) or (10). %\eqref{afm_sc} or \eqref{fm_sc}.
We treat the amplitude of the SC OP, $|\Delta|$, as a free parameter. For fixed filling $\nc$, we then show the single-particle gap $E$ of the BdG band structure as a function of the amplitude $|\Delta|$. In three of the four plots shown, the gap closes at least once for $|\Delta|/t > 10^{-1}$. We also calculate the Chern numbers for each of these gapped phases using numerical integration\,\cite{fukui2007}, and see that the Chern number often changes between these gapped phases. 
Since the critical scales of the FRG divergence into the superconducting phases are all $\Lambda_c < 10^{-1}$, Fig.\,1,  we assume the phase with $|\Delta|/t \lesssim 10^{-1}$ to be the physical one. This gives the Chern numbers for the superconducting phases as reported in the main text without the need for a fully self-consistent treatment.

\section{Ribbon spectra}

\begin{figure}[h!]
    \centering
    \includegraphics{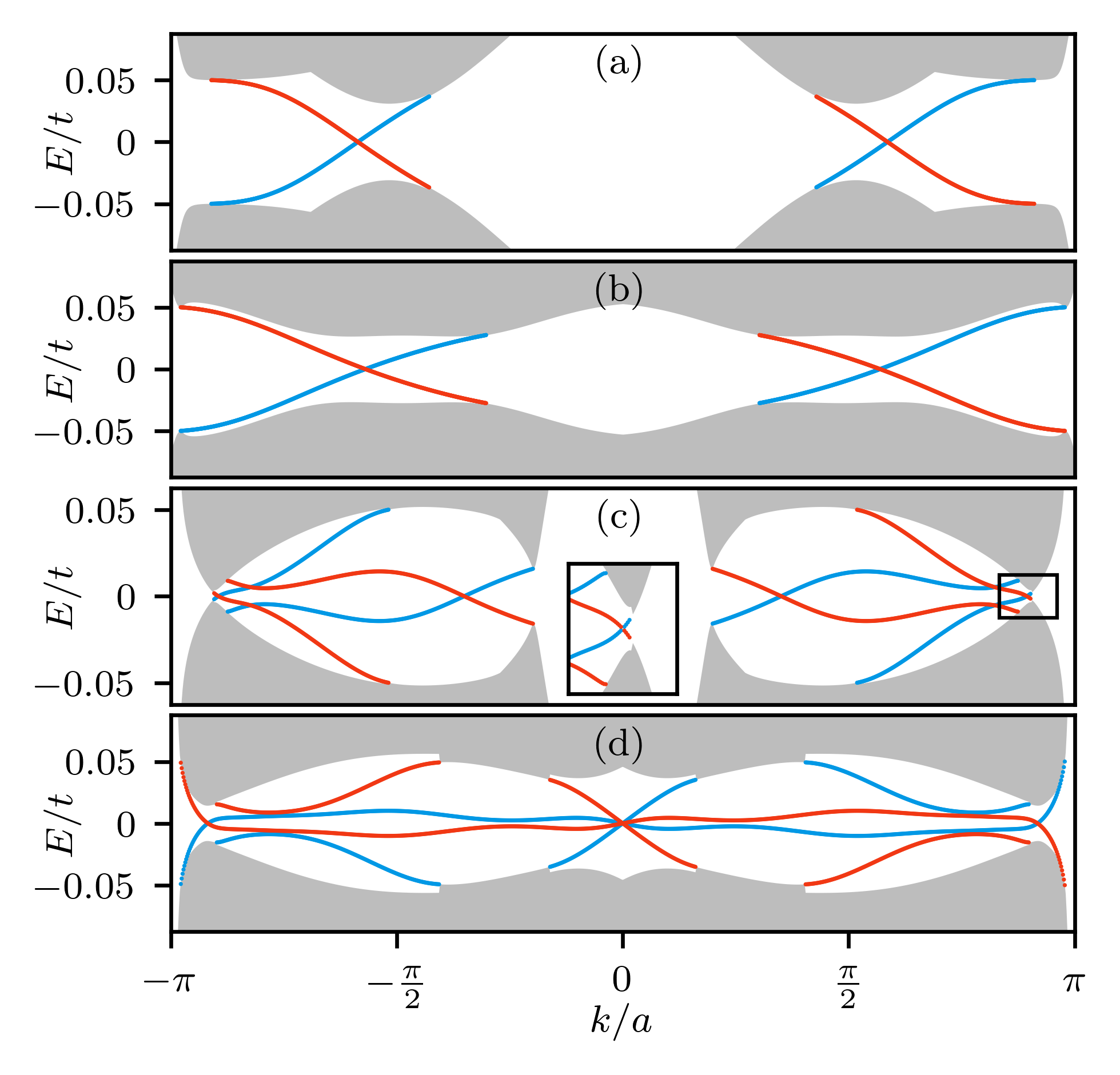}
    \caption{Single-particle spectra of superconducting states. Bulk bands are shown in grey, gaps in white and the chiral edge modes on the left (right) edge in red (blue). Four representative parameters are chosen for the phases above and below van-Hove filling for AFM and FM cases.
(a) AFM case, $\nc=1.18$. (b) AFM case, $\nc=1.30$. (c) FM case, $\nc=1.18$. (d) FM case, $\nc=1.30$.
For all plots, $U/t=3.0$ and $|\Delta| = 0.1$. Note that the chiral edge modes in (a) and (b) are 
spin-degenerate.
}
    \label{fig:ribbons}
\end{figure}

In addition to a calculation of the Chern number by numerical integration of the Berry curvature over the Brillouin zone\,\cite{fukui2007}, the Chern number can be calculated by Laughlin's edge counting for states with a ribbon geometry\,\cite{laughlin1981}. The expression of the SC order parameter (OP) in terms of form factors means that TUFRG is well suited to calculate ribbon spectra, and the methodology of such a setup is discussed in a previous work\,\cite{bunney2024}.

The ribbon spectrum is calculated from the BdG matrix in the ribbon geometry, which takes the chiral superpositions of the SC OP, \textit{i.e.,} the output of the FRG divergence analysis, as input.
We set the SC OP amplitude to be $|\Delta| = 0.1$, which is related to the components $\Delta_{\sigma\sigma'}(\boldsymbol k)$ via
\begin{equation}
    |\Delta|^2 = \frac{1}{4N} \sum_{\boldsymbol k \, \sigma \sigma^\prime} |\Delta_{\sigma \sigma^\prime} (\boldsymbol k)|^2
\end{equation}
where we sum over the $N$ discrete momentum points $\boldsymbol k$. 
By choosing a value for $|\Delta|$ the amplitude of the entire SC OP is fixed.
The chosen $|\Delta|$ is large enough to resolve the SC gap in the ribbon spectra, see Fig.\,\ref{fig:ribbons}, whilst remaining in the same topological phase as the expected physical $|\Delta| \sim \Lambda_c$\,\cite{platt2013}.

Edge states are identified where more than $90\%$ of the spectral weight of the BdG eigenvector lies on one half of the ribbon, and the Chern number is equal to the net number of the left or right states which traverse the gap in the one direction, \textit{e.g.,} from the lower bands to the upper.

In Fig.\,\ref{fig:ribbons} we show the ribbon spectra for representative parameters -- one below and one above van-Hove filling for both AFM and FM Hubbard--Kitaev model at $U/t=3$ and $|\Delta|=0.1$. The number of chiral edge modes per edge matches the Chern number calculations, see main text.

\section{Relation to and distinction from previously reported superconducting states in Kitaev-based models}

Superconductivity in models derived from the Kitaev spin liquid has been discussed in earlier work, and we summarize similarities and differences here.

The Kitaev-Kondo lattice and its FL$^\ast$-FL transition have been studied using Majorana-based mean-field theory in Refs.~\onlinecite{seifert2018,choi2018}. Both papers reported superconductivity on the FL side of this transition, i.e., at intermediate Kondo coupling, with slightly different superconducting properties due to different mean-field schemes. The superconducting states are topological in the sense of non-trivial band invariants. They are, however, not fractionalized and do not display an emergent gauge structure, as they derive from a conventional (heavy) Fermi-liquid parent state; these are respresented by the phase labelled SC in Fig.~1 of the main paper. Technically, the onset of Kondo screening eliminates the internal gauge structure via a Higgs-type mechanism and leads to confinement.
The model has also been studied in Ref.~\onlinecite{lundemo2024}, there at weak coupling. Not unlike the present work, these authors derived an effective model for the conduction electrons. Their results are restricted to FM Kitaev coupling and small doping away from half-filling and appear broadly consistent with ours.

A different route to superconductivity has been explored in Ref.~\onlinecite{you2012}: This work studied a single-band $t$-$J$-type model, describing a Kitaev spin liquid when doped away from half filling. The paper again employs Majorana-based mean-field theory and reports two superconducting phases as function of doping level, dubbed SC1 and SC2. While SC2 derives from a Fermi-liquid normal state and is BCS-like, SC1 appears more complicated. It descends from a fractionalized normal state, with spinons and holons, and inherits some of the projective-symmetry-group structure from the Kitaev spin liquid. We believe, however, that the holon condensation which drives the superconducting transition again eliminates the gauge structure via Higgs, as the holon carries a gauge charge. Then, SC1 is a non-fractionalized superconducting state. This underlines the difficulty to obtain a fractionalized SC$^\ast$ phase using mean-field-type approaches to the original Kitaev interactions.

%\newpage

%%%%%%%%%%%%%%%%%%%%%%%%%%%%%%%%%%%%%%%%%%%%%%%%%%%%%%%%%%%%%%%%%%%%%%%
%%%%%%%%%%%%%%%%%%%%%%%%%%%%%%%%%%%%%%%%%%%%%%%%%%%%%%%%%%%%%%%%%%%%%%%
%%%%%%%%%%%%%%%%%%%%%%%%%%%%%%%%%%%%%%%%%%%%%%%%%%%%%%%%%%%%%%%%%%%%%%%

\bibliography{kitaev_bib}